# Modernized Latvian Ergonomic Keyboard

dr. Comp. Sci. Valdis Vitolins

Odo Ltd, Latvia

**ABSTRACT**

Increasingly more people use computers and create content using keyboards (even with leading edge touch-screen technology). As in the most part of the world, in Latvia also conventional "Qwerty" keyboard is used. Though for Latvian it is much worse than for English, especially due to enormous load to little fingers. It causes repetitive strain injuries and affects productivity of workers with extensive keyboard usage, especially for data input operators, call centers, inquiry office workers, etc. Improving computer keyboard layout decrease stress to hands and fingers thus minimizing exhaustion and injuries.

With analysis of English and Latvian public domain novels and modern texts, letter appearance an sequence distribution for Latvian language was found. Qualities of alternative layouts for English (Dvorak, Colemak, Hallinstad) were investigated and open source carpalx simulation tool was adjusted according to the findings. Then carpalx was used to check more than 25 million keyboard layouts, measuring finger/hand effort, stroke typing convenience etc., to find the best one. It was proved that existing "Šusildatec" (classic Latvian Ergonomic standard) keyboard is only slightly better than "Qwerty" for Latvian, though it is much worse for English.

After computer simulation, several best layouts were tried practically for more than 6 months and most convenient one was promoted as a new "Latvian Modern" keyboard. Its typing effort is less than for "Šusildatec", load is distributed according to finger strength, and typing strokes are alternating better between hands and fingers. Comparing to "Qwerty" keyboard new layout is better not only for Latvian but for English also. Keyboard drivers are developed for Microsoft Windows and Linux operating systems and are freely available in the web under permissive license.

**KEYWORDS**

Latvian, ergonomic, keyboard, layout, typing, effort, simulation

# Introduction

Distribution of characters (letters and punctuation marks) is not even and few characters appears much more recently than others. Although precise distribution of characters differs due to author's style, protagonist's names and paper domain, any sample in general conforms to the English language [1]. Fig 1. shows count of characters in "The Adventures of Tom Sawyer" by Mark Twain.

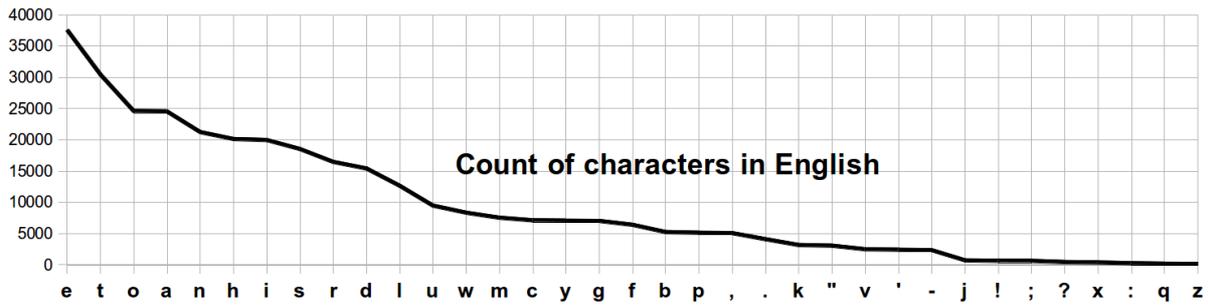

*Fig. 1: Count of characters in "The Adventures of Tom Sawyer"*

People who's native language is not English, incorrectly assume that the most popular Qwerty keyboard layout is adapted for English language. Actually it is wrong assumption, because *it was developed considering only marketing and jamming of typewriter hammers* [2]. To avoid jamming, most frequent letters were evenly spread around the keyboard. By looking at worn keyboard one can see picture similar to (Fig. 2), where wearing is shown as a gradual shading of keys, applied in relation to letter appearance in English. Home position for fingers are shown with black rectangle. Positions of top five most common letters are shown with numbers.

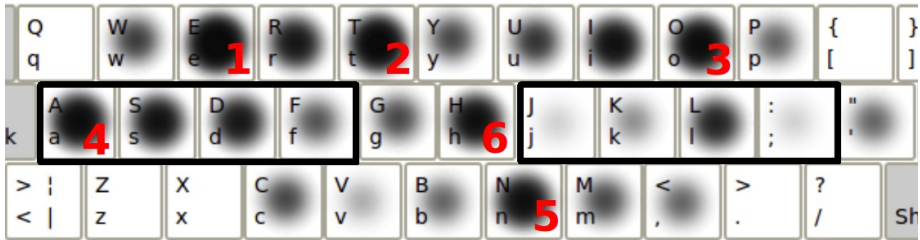

*Fig. 2: Qwerty keyboard "wearing" for English language*

Inconveniences of Qwerty keyboard are widely analyzed and several layouts are provided: Dvorak, Colemak, etc. [3, 4, 5]. One particular feature of Qwerty keyboard for English language is that although fingers regularly jump to upper keyboard row (three of the most common letters are on the top row), finger effort is distributed quite appropriately — stronger (index, middle) fingers are used more frequently than weaker (ring, little) ones.

Distribution of letters in Latvian differs from English. By summarizing different texts in Latvian, appearance of characters is shown in Fig. 3. Notable difference for Latvian is that accented characters (long vowels: ā, ē, ī, ū, soft consonants: š, ķ, etc.) on Qwerty keyboard are typed using so called "dead key" before plain (i.e. unaccented) letter. Thus *for standard Qwerty keyboard the third most used key in Latvian is not letter but dead key* (usually Apostrophe):

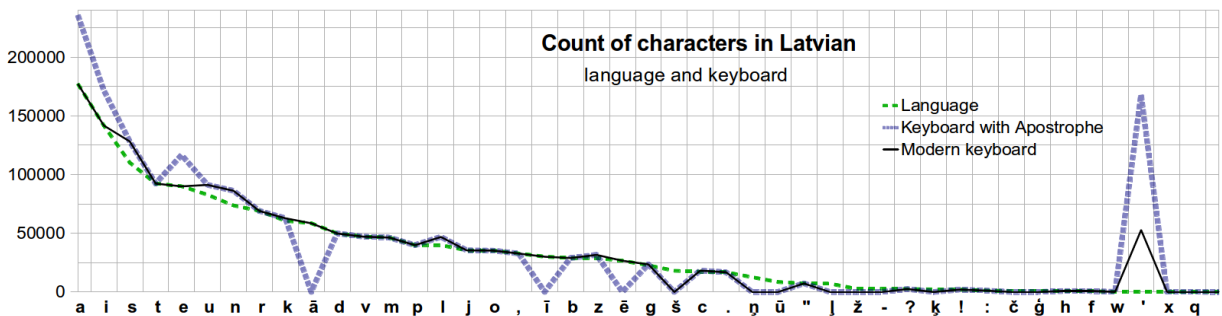

*Fig. 3: Appearance of characters in Latvian language and keyboards*

Looking at character distribution on Qwerty keyboard layout (Fig. 4) one can see that too much load is distributed for left hand (letter A) and right hand little fingers (Apostrophe).

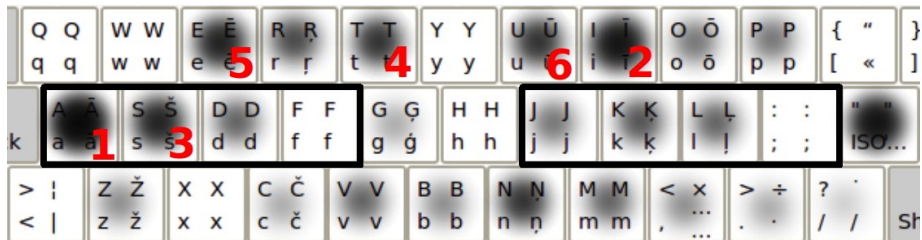

*Fig. 4: Qwerty keyboard "wearing" for Latvian language*

In typewriter times *Latvian Ergonomic* keyboard (also called Šusildatec or Ūgjrm) was developed for Latvian language. Notable feature of the layout is that most common letters are under index fingers and further letters are distributed to outer fingers. Though 7[th] and 8[th] most common letters (N and R) are not placed under little fingers but in index finger's upper row, because for mechanical typewriters little finger was too weak for regular use. This layout has separate key for all Latvian letters sacrificing several Latin letters and special characters (see Fig. 5):

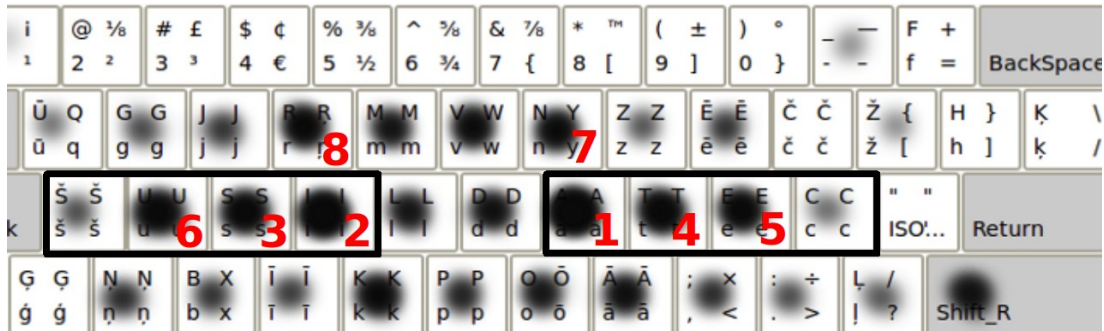

*Fig. 5 Latvian ergonomic keyboard "wearing" for Latvian language*

Although Latvian ergonomic keyboard is more convenient than Qwerty keyboard for Latvian, it has several drawbacks:
1. Several accented letters (Š, Ģ, Ū, Ž, Ķ, Ņ) are placed at inconvenient positions which are not related to their unaccented letters (S, G, U, Z, K, N). Many accented letters are placed under other hand without any system. That makes typing hard to learn.
2. Even index fingers are the most durable, they are used *too* frequently, because under them are placed most common letters (A and I), and these fingers also press twice as much keys (8, including number row) as other fingers (4). So, most of type speed depends only on agility of index fingers.
3. Several common letters are located in lower row, which is considered less convenient than upper row.
4. Several Latin letters (Y, W, Q, X) can be typed only invoking Alt key. Though due to globalization they deserve more prominent place, e.g.: *www*, *Linux*, *yes* are much more common in every day use than Latvian: *ģeorģīne*, *ļurļaks* or *ņuņņa*. Also special characters (e.g. slash and column) are mandatory part of web address but are hard to type.
5. Commonly used shortcut keys are not considered. E.g. *Ctrl+X* shortcut key needs also Alt key, because X letter can be get only in such way. Commonly used X, C, Z, V letters are not placed nearby to each other.

So, *Latvian ergonomic keyboard was appropriate for typewriters, but it is not convenient enough for contemporary computer usage*. Although formally it was accepted as a National standard [6], it is not widely accepted for computers (in difference with Russian ЙЦУКЕН keyboard which layout is developed by very similar approach).

**Development of Latvian modern keyboard**

Development of Latvian modern keyboard was long and graduate process, which started with frustration on Qwerty keyboard and (quite surprising) discovery that so called "Latvian ergonomic" keyboard is not much simpler/convenient than Qwerty. From the beginning, investigation had purely practical reasons, but it was advanced to even deeper analysis with more sophisticated and precise methods.

To minimize systematic errors in letter distribution as a language reference were used different publicly available sources: "[Mērnieku laiki](#)", "[Ceplis](#)", "[Purva bridējs](#)", "[Ugunszīme](#)" and personal diary (together more than 2 million characters). Distribution of letters and words were got using Linux text processing tools such as: [grep](#), [sed](#), [awk](#) and [wc](#). First attempt was to adjust Latvian ergonomic keyboard, but it was abandoned because any small adjustments were not sufficient. Next evolution was achieved using Java application [7], with manual layout adjustment counting total distance what fingers travel over keyboard, relative time of fingers is keyboard rows, load distribution for hands and fingers, typing alternation between hands and fingers. As most effective layout was found wit USIRVPNATEY letters in the home row (see *usirvp* in Fig. 7).

Practically using this layout, it was found that important aspect of usability is convenient layout of commonly used hot keys: X, Z, C, V (invoked together with Ctrl key). They were placed too far away each to other. During investigation of public sources, another tool carpalx [8] was found. This tool does optimization himself by random changes in keyboard layout and checking performance. Performance is based on triads, which are three character substrings formed from the text. The effort model takes into account contributions of following main characteristics:
1. finger travel distance over the keyboard (base effort) if fingers are moved less, then they travel less.
2. uneven hand usage penalty,
3. weaker finger usage and not-home-row usage penalties,
4. typing stroke penalty (e.g. QXE when hand move up and down or H and ' for horizontal movement).

Important aspect is that several usability features conflict each to other. E.g. by improving layout for Latvian, in general it becomes worse for English. Decreasing load on little fingers makes worse stroke typing experience, as it increases possibility that the same finger will be used again in different position. Accented and unaccented letters and hot keys sticked together decrease key swapping possibility between hands, etc.

Searching for optimized keyboard layouts was resource consuming task. Three computers were used for several days and were checked more than 25 million layouts. Even this is huge number, it is small comparing to $10^{59}$ possible combinations. Therefore it can't be granted that the best possible layout is found, though results allow to assume, that it should not be better than 15% for chosen effort parameters.

Although typing effort model in carpalx is quite sophisticated, it doesn't cover all aspects of ergonomics. E.g., carpalx doesn't deal with keyboard "aesthetics" and doesn't check if it is possible to group accented letters together with unaccented ones. It also doesn't check finger effort, which is not related to writing (e.g. common key sequences and combinations in application usage). Therefore layouts found by carpalx where manually "tweaked" and checked, how much it can be improved considering these constraints without severe decrease in typing effort. Fig 6. shows comparison of Qwerty (asdfg), Latvian ergonomic (susild) and other keyboards for total typing effort (summing all efforts and penalties) for English and Latvian languages. Keyboards which were practically used for longer time are shown with bigger marks (usirvp, euankd and euank') and as final one euank' layout was chosen. It can be seen that in place of smallest effort little bit worse is chosen due to other considerations which can't be checked by carpalx tool.

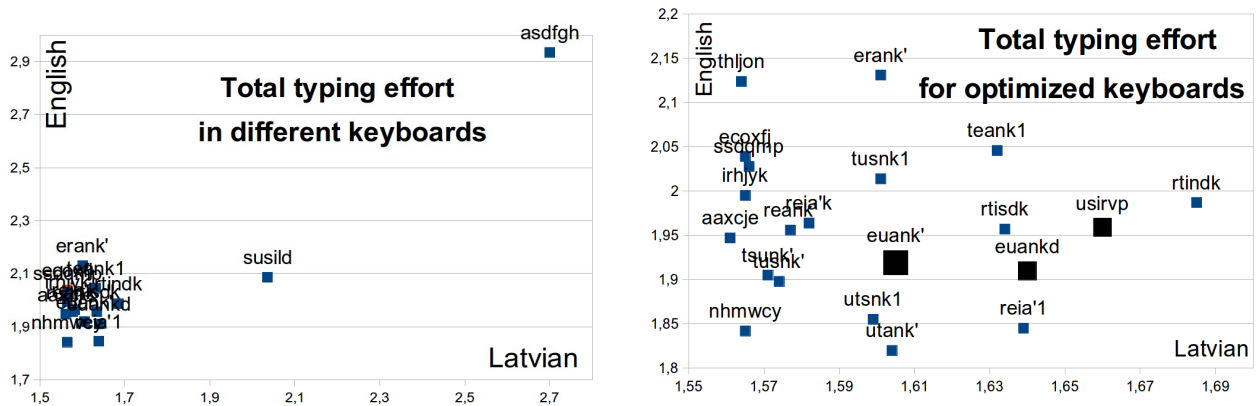

*Fig. 6: Total typing effort for different keyboard layouts*

## Results

To ensure that keyboard layouts found by carpalx as most ergonomic, were practically usable in real life they were tested practically. Keyboard drivers were prepared for Linux X.org Xwindow system (xkb configuration file) and Microsoft Windows operating systems using [MS Keyboard Layout Creator](). For Linux operating system are developed typing training lessons in [Ktouch]() tool (can be run also in Windows, using ported KDE version).

In time for more than six months, three keyboard layouts were practically tested for more than three weeks, dosen of layouts were tried for one day. Using practical experience, besides letters in keyboard were added additional symbols (copyright ©, parentheses: «»""", slashes and pipe: /|\, smilies: ☺, etc.) Fig 7. shows latest version of the keyboard layout in Ubuntu Linux operating system.

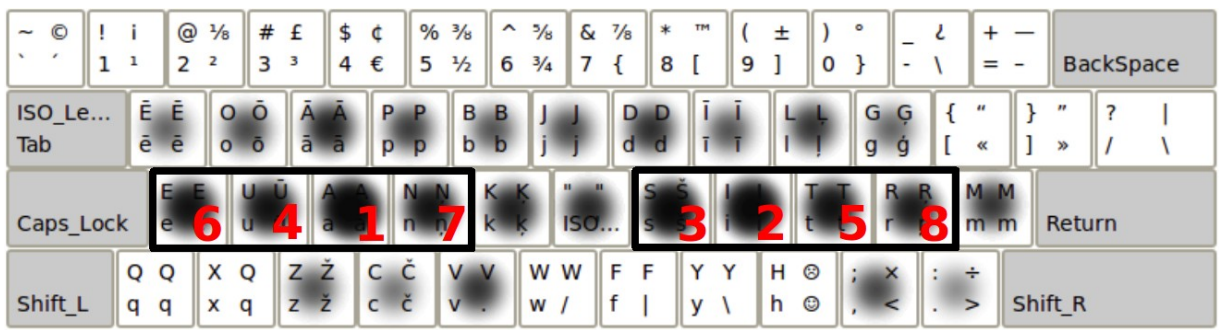

*Fig. 7: Latvian modern keyboard "wearing" for Latvian language*

From the web forum it is known that at least *two* other people have started using this keyboard and are very happy with results. Even though it is *not much* it shows that there is interest in this layout. One of these enthusiasts has included this layout in the newest X.org package.

**Benefits of the Latvian modern keyboard**

*In comparison to Qwerty keyboard*
1. 8 most popular letters are located under fingers and can be typed without movement to other key. These letters make more than half of the all amount of letters.
2. With modern layout can be typed at least 15 times more words without moving fingers than with Qwerty keyboard.
3. Introducing dedicated keys for most popular accented letters (Ā, Ē, Ī) number of pressed keys is decreased by 15%.
4. Total path what fingers travel over keyboard is decreased ~1,5 times not only for Latvian, but also for English language.
5. About 95% of letters can be typed by only single key pressing (i.e. one letter is typed by one key), other 5% can be typed by two key stroke, which is more convenient than for Qwerty layout.

*In comparison to Latvian ergonomic (Šusildatec) keyboard*
1. Even though in modern layout should be pressed 2% more keys (i.e. Alt or Apostrophe as dead key before missing accented letters), in summary fingers travel over keyboard by10% less.
2. By moving most common letter position from index finger to middle finger, probability of typing next letter with the same finger is decreased 3 times.
3. Fingers are located 2 times more in their home position, and are 2 times less moved to the lower row.
4. With modern layout can be written at least 2 times more words without moving fingers than with Šusildatec keyboard.
5. By eliminating rarely used diacritic letters Ņ, Č, Ģ, Ņ, Ķ, Ļ and Ž, place is cleared for Latin letters and special marks. This greatly improves typing in English and application usage (browsing web, reading and writing e-mail, usage of system utilities, programming etc.).
6. Most frequently used shortcut keys X, Z, C, V are placed together.
7. Diacritic letters Ā, Ē, Ī are located above their "plain" letters, all Latin letters which are not used in Latvian are in lower row, what greatly improves learning.

*Drawbacks of the Latvian modern keyboard*
1. Several diacritic letters: Ņ, Ū, Ļ, Ķ, Ž, Ģ and Č (and in Latvian standard not used Ŗ and Ō) are typed by two keys using before pressed dead key (Apostrophe) or simultaneously pressed Alt key. Such letters are less than 10% from all.
2. Apostrophe itself can be typed pressing this key twice ore pressing this key and space key (similarly to Qwerty keyboard).
3. Backslash \ less/greater than <> and pipe | is typed by dead key before or with simultaneously pressed Alt key.
4. Keyboard is not fully optimized for Latvian allowing small (<10%) typing effort increase:
    1. Common hot keys X, Z, C, V are grouped together to improve application usability,
    2. To improve typing for English, hot keys follows as XZCV, only for Latvian XCZV would be better.
    3. For English upper row key sequence is ĒOĀPBJDĪLG, only for Latvian ĒPĀDBGJĪLO would be better.
5. In compact keyboards (e.g. on laptops) letter Q doesn't have his own key. Then it can be typed by dead key before or with simultaneously pressed Alt and X key.
6. Apostrophe/double quotes key is between letters, what could look surprisingly/funny.

## Conclusion and future work

Investigations in computer-human interaction with speech-to-text tools is going, though they are not reliable enough for everyday use, and don't seem feasible for busy call center or other open-space office. So, keyboards will be used long time further and it is not too late to implement new standard layout for computer keyboards in Latvia similarly to Russian ergonomic ("ЙЦУКЕН") layout. Although the most improvement the new layout could provide for intensive typists, increasingly more people have personal computers and they are free to use this layout on computers they own.

All investigation is performed by freely available open source tools, process is documented [9] and results are published [10] in the web under permissive (Creative Commons — Attribution) license. This allows anybody to reuse achieved results and perform further investigation. If somebody wander is it worth to find out other keyboard layout, for his language, he can perform simple qualitative analysis, by checking following keyboard features:

1. If two most frequent letters in language (e.g. E and T in English) do not appear under fingers in their home position (e.g. ASDF JKL; for Qwerty keyboard), keyboard is very far from ergonomic.
2. If most frequent two letters appear under index fingers (e.g. A and I in Latvian or A and O for Russian ergonomic keyboards), highly probably index fingers are used too much and layout can be improved by distributing load to other fingers.
3. If most frequent letters appear under middle fingers, (e.g. in E and T for Dvorak, QGMLWY and QFMLWY in English, A and I for Modern keyboard in Latvian), keyboard seems to be appropriate to ergonomics standards.
4. If all 8 most frequent letter appear under fingers in their home position (e.g. Hallingstad for English, Modern keyboard for Latvian), keyboard has smallest finger traveling distance (base effort). Though it makes bigger load to little fingers, or sacrifice stroke typing convenience in comparison to layouts with slightly bigger finger travel effort.

**REFERENCES**
1. *Letter frequency*, http://en.wikipedia.org/wiki/Letter_frequency
2. **Weller, Charles Edward** (1918), The early history of the typewriter, La Porte, Indiana: Chase & Shepard, http://www.archive.org/details/earlyhistorytyp00wellgoog
3. **Piepgrass**, **David**, Why QWERTY, And What's Better? http://millikeys.sourceforge.net/misc/why-qwerty.pdf
4. **Krzywinski, Martin**. "Colemak - Popular Alternative". Carpalx - keyboard layout optimizer. Canada's Michael Smith Genome Sciences Centre: http://mkweb.bcgsc.ca/carpalx/?colemak
5. **Ober, Scoth**, Relative Efficiencies of the Standard and Dvorak Simplified Keyboards.
6. Valsts standarts, Informācijas tehnoloģija, Datoru tastatūras, Latviešu tastatūra datoriem (LVS 23-93)
7. *Keyboard comparison applet*, http://colemak.com/Compare
8. **Krzywinski, Martin**, Carpalx keyboard effort simulation tool, http://mkweb.bcgsc.ca/carpalx/
9. **Vītoliņš, Valdis**, Modernizēta latviešu ergonomiskā tastatūra, http://odo.lv/LatvianKeyboard
10. **Vītoliņš, Valdis**, Ergonomiskās tastatūras pievienošana Windows un Linux operētājsistēmās: http://odo.lv/Recipes/LatvianKeyboard